\documentclass[twoside]{article}
\usepackage{qic,epsfig}
\usepackage{amsmath}
\usepackage{openbib}
\usepackage{notoccite} % looking for this for a long time, it orders references by citation
\usepackage{subeqnarray}
\usepackage{graphicx}
\usepackage{wrapfig}
\usepackage{floatrow}
\input{Qcircuit}

\newfloatcommand{capbtabbox}{table}[][] % for the table side by side with figure
\newtheorem{myconj*}{Conjecture} % star so not numbered

\begin{document}
\setlength{\textheight}{8.0truein}    %FOR 2ND PAGE ONWARDS

\runninghead{Quantum circuits for simple periodic functions  $\ldots$}
            {Gamel and James $\ldots$}

\normalsize\textlineskip
\thispagestyle{empty}
\setcounter{page}{1}

\newcommand{\eq}[1]{{eq.(\ref{#1})}}
\renewcommand{\theequation}{{\arabic{section}.\arabic{equation}}}  %%%  e.g. (4.12) %%
%\DeclarePairedDelimiter{\ceil}{\lceil}{\rceil}

%\copyrightheading{Vol.}{No.}{Year}{Page Nos.}
\copyrightheading{0}{0}{2013}{000--000}

\vspace*{0.88truein}

\alphfootnote

\fpage{1}

\centerline{\bf
%%%%%%%%%%%%%%%%%%%%%
%Put in titiles here
%%%%%%%%%%%%%%%%%%%%% Synthesizing Quantum Circuits for Simple Periodic Functions
SYNTHESIZING QUANTUM CIRCUITS FOR SIMPLE PERIODIC FUNCTIONS}
\vspace*{0.035truein}
\centerline{\bf FOR QUANTUM INFORMATION AND COMPUTATION} %\footnote{Typeset the
%title in 10 pt Times Roman, uppercase and boldface.}}
\vspace*{0.37truein}
\centerline{\footnotesize
%%%%%%%%%%%%%%%%%%%%%%%%%%%%%%%%%%%%
%put authors' name and address here
%%%%%%%%%%%%%%%%%%%%%%%%%%%%%%%%%%%%
OMAR GAMEL %\footnote{Typeset names in
}
\vspace*{0.015truein}
\centerline{\footnotesize\it  Department of Physics, University of Toronto, 60 St. George St.,}
\baselineskip=10pt
\centerline{\footnotesize\it Toronto, Ontario, M5S 1A7, Canada}
%Country\footnote{State completely without abbreviations, the
%affiliation and mailing address, including country. Typeset in 8
%pt Times Italic.}}
\vspace*{10pt}
\centerline{\footnotesize 
DANIEL F. V. JAMES}
\vspace*{0.015truein}
\centerline{\footnotesize\it  Department of Physics, University of Toronto, 60 St. George St.,}
\baselineskip=10pt
\centerline{\footnotesize\it Toronto, Ontario, M5S 1A7, Canada}
\vspace*{0.225truein}
\publisher{(May 15, 2013)}{(October 5, 2013)}
%\publisher{(received date)}{(revised date)}

\vspace*{0.21truein}

%% \abstracts{first paragraph}{second paragraph}{third paragraph}
%% If there is only one paragraph, just keep the second and third empty 
%% like the following one 
\abstracts{
%%%%%%%%%%%%%%%%%%%%
% put abstract here
%%%%%%%%%%%%%%%%%%%%
Periodic functions are of special importance in quantum computing, particularly in applications of Shor's algorithm. We explore methods of creating circuits for periodic functions to better understand their properties. We introduce a method for constructing the circuit for a simple monoperiodic function, that is one-to-one within a single period, of a given period $p$. We conjecture that to create a simple periodic function of period $p$, where $p$  is an $n$-bit number, one needs at most $n$ Toffoli gates. 
}{}{}

\vspace*{10pt}

\keywords{Quantum Circit, Periodic Function, Shor's Algorithm}
\vspace*{3pt}
\communicate{to be filled by the Editorial}

\vspace*{1pt}\textlineskip    %) USE THIS MEASUREMENT WHEN THERE IS
   %) A SECTION HEADING
%\vspace*{-0.5pt}
%\noindent
%%%%%%%%%%%%%%%%%%%%%%%%%%%%%%%%
%put the text of the paper here
%%%%%%%%%%%%%%%%%%%%%%%%%%%%%%%%
\section{Introduction}
% clarify intro, this is not factoring, but creating periodic state, proof of concept - clarify what exactly we are doing and which parts of Shor's are being demonstrated
As quantum computers improve and grow in size beyond a dozen or so qubits, they face two daunting problems. Firstly, since complete tomographic reconstruction of the quantum state becomes increasingly intractable \cite{flammia, jamesKaz}, how might these devices be characterized and their performance validated? Secondly, and by no means distinctly, is the problem of finding meaningful milestones for device development along the long road to true large scale devices capable of tackling useful problems.

Shor's factoring algorithm \cite{shor}, nearly 20 years since it was discovered, remains arguably the most promising and compelling application of quantum computing. It allows one to factor large numbers in polynomial time, undermining the most common cryptographic schemes in use today, such as RSA cryptography \cite{rsa}. Highly simplified versions of this algorithm have been implemented using both NMR techniques \cite{vandersypen} and linear optical implementations \cite{lanyon,lu,martin}, providing just such a milestone for two of the possible technologies under consideration as quantum computers. 

Shor's algorithm makes use of the {\em periodicity} of the modular exponential function, whose period can be evaluated efficiently due to the inherent massive parallelism of quantum computing. The quantum Fourier transform can then be used to extract the period of the function, from which desired factors can be deduced. Periodicity is also of particular interest, since it has been shown that it is the size of the associated period rather than the number being factored which determines the difficulty of the factorization \cite{jsmolin}. This motivates us to better understand periodic functions as an independent object of study. In particular, it is of interest to study how a simple quantum circuit can be created to implement functions of a given period, and the resources such a circuit requires. Moreover, given the limited capacity of currently realizable experimental systems, it is of interest to find the minimal number of gates needed to create the circuit for a function of a given period. 

Synthesizing these simple circuits will provide interesting milestones as experimental technology advances. Furthermore, periodic functions are relatively straightforward to verify, via a quantum Fourier transform \cite{griffithsniu}. Therefore, they may play a role in validating new quantum devices as they are introduced. That is, we use a new device or technology to implement a periodic quantum circuit, and then use the quantum Fourier transform to check that the circuit indeed produces a periodic superposition state when it should, to ensure that the new device functions correctly.

There has been much literature published on the synthesis of quantum circuits in the context of reversible computing theory, using Toffoli gates \cite{maslov, saeedi3, gupta, wille}. However, the algorithms they presented are fairly general, whereas we intend to focus on simple periodic functions. 

In this paper, we investigate the process of creating a quantum circuit for a simple periodic function of a given period $p$, using only the basic quantum gates: CNOT and Toffoli gates \cite{toffoli}. Note that Toffoli gates along with local unitary gates, form a universal set of gates \cite{toffoli}. We begin in section \ref{secperiodicfunc} by defining the types of periodic functions and the class of simple periodic function we choose to investigate. In section \ref{sec3circuitsyn} we explain the process of synthesizing some circuits of the simple periodic functions, while giving some illustrative examples for some values of period $p$. More circuits are provided in the appendix. Finally, in section \ref{sec4reqres} we list the required resources (gates) found in synthesizing the circuits for up to 5-bit periods $p$. In doing so, we conjecture an upper limit to the required number of Toffoli gates in such a circuit for any period $p$. 

\section{Periodic Functions}\label{secperiodicfunc}

We define a \emph{periodic injective} function as any function $F_{p,n}$ satisfying the following properties: 

\begin{enumerate}
\item It is a binary function with $n$ input bits and $m$ output bits. The input $x$ is an integer satisfying $0 \le x < 2^n$, and output $y$ satisfies  $0 \le y < 2^m$.
\item It has a period $p$, i.e. $F_{p,n}(x) = F_{p,n}(x-p)$ for $x \ge p$.
\item It is one-to-one (injective) within a single period, i.e. $F_{p,n}(x_1) = F_{p,n}(x_2)$ for $0 \le x_1, x_2 < p$ implies $x_1=x_2$. 
\end{enumerate}

For example, looking at continuous functions instead of discrete ones for the moment, we see that $\tan$ has period $\pi$ and is injective within a single period. While $\sin$ and $\cos$ both have period $2\pi$ and are not injective within a period. 

Writing the input variable $x$ in binary notation, we have $x= x_n ... x_2 x_1$, where the $x_i$ denote each of the $n$ input bits. For example, if $x$ has the value $13$ in decimal notation, then in binary notation it is $x=1101$, with individual bits $x_4 = x_3 = x_1 = 1$ and $x_2=0$. Similarly, if $y$ is the output, we write it in binary notation as $y = y_m ... y_2 y_1$ and the $y_j$ denote each of the $m$ output bits. Note that $m$ need only be large enough so the number $p$ can be represented in $m$ bits, i.e. $m= \lceil \log_2(p) \rceil$, where $\lceil w\rceil$ is the ceiling of $w$ (defined as the smallest integer which is not smaller than $w$; for example $\lceil 3.142 \rceil = 4$, $\lceil 5 \rceil = 5$).

We further define a \emph{monoperiodic} function $G_p$ as a periodic injective function with period $p$, and a number of input bits enough to contain just one complete period $p$. That is, $n=m= \lceil \log_2(p) \rceil$. So we write
\begin{equation}
G_p \equiv F_{p, \lceil \log_2(p) \rceil}.
\end{equation}
% would it be better to define these as sets instead?

Note that there are many different monoperiodic functions $G_p$ for a given $p$. All of them have the same basic structure, and are related to each other through a simple isomorphism on integers in the output space (i.e. a straight forward relabeling of output states). For example, in the period 3 case, suppose we define $V_3(x)\equiv x \mod 3$, and $W_3$ is defined such that $W_3(0)=W_3(3)=1, W_3(1)=0$, and $W_3(2)=2$, where the argument satisfies $0 \le x < 4$. Then both $V_3$ and $W_3$ qualify as monoperiodic functions of period 3, and they can be related via the isomorphism $t$ defined as $t[0]=1, t[1]=0, t[2]=2,$ and $t[3]=3$, satisfying $W_3(x)=t[V_3(x)]$. This example isomorphism $t$ happens to be self-inverse, but that need not always be the case. The fact that they can be related via isomorphisms means all monoperiodic functions of a given period are linked by a type of equivalence relation.

However, despite this mathematical equivalence relation between all monoperiodic functions of a given period $p$, there is some practical difference between them. Suppose we wish to synthesize a quantum circuit for each of these functions using only the basic quantum gates, CNOT and Toffoli gates. It turns out that some of them will be easier to construct than others, requiring fewer gates. 

In what follows, we seek to synthesize the quantum circuit for the monoperiodic function of period $p$ that is easiest to implement. We define the easiest function as one with the smallest gate count. Toffoli gates are the most demanding to implement experimentally. If one implements a Toffoli gate via CNOT gates on a system of qubits, it takes six of the latter to implement, in addition to some comparatively cheap single qubit gates \cite{barenco,shende,toffoliralph}. It takes fewer resources if we allow qutrits in an optical implementation \cite{lanyon2} or use vibrational modes in an ion trap to store additional information \cite{blatt}, however for our purposes we will count a Toffoli gate as equivalent to six CNOT gates. 

There are many methods to quantify the cost of synthesizing a quantum circuit \cite{wille}. We define $N_{T}$ and $N_{CN}$ as the number of Toffoli and CNOT gates for a given circuit, respectively. We will focus on Toffoli gates since they are the main drain on resources, and minimizing their number is of most interest. We will also pay attention to the \emph{quantum cost}, defined as $Q \equiv N_{CN} + 6N_{T}$. The quantum cost is roughly equivalent to the number of elementary two-qubit (CNOT) gates needed to implement a circuit.

Suppose we implement the most efficient quantum circuit for all possible monoperiodic functions $G_p$. We define a \emph{simple periodic function} $S_p$ as one which minimizes the Toffoli gate count $N_{T}$ among all monoperiodic functions $G_p$. Again, $S_p$ may or may not be unique, as there may be multiple functions with this minimum Toffoli gate count. 

To summarize, there are many periodic injective functions with period $p$, denoted $F_{n,p}$. A subset of them, monoperiodic functions denoted $G_p$, has only one complete period in the domain of its input qubits. A smaller subset of these, simple periodic functions denoted $S_p$, minimizes the Toffoli gate count during the circuit synthesis. It is this final subset we are interested in.

\section{Circuit Synthesis}\label{sec3circuitsyn}

We now address the task of synthesizing the quantum circuit for a function $S_p$, for a given $p$. In fact, it is by synthesizing the circuit for a monoperiodic function $G_p$ while trying to minimize the number of Toffoli gates that $S_p$ can be found. We conjecture that the process used below to synthesize the circuits minimizes the number of Toffoli gates. A rigorous proof will be addressed in a future work.

Note that if $p$ is even, then the quantum circuit for $S_p$ is simply the circuit of $S_{\frac{p}{2}}$ synthesized between the input and output bits $x_i$ and $y_i$ for ($i=2,...,n$), with an additional CNOT gate that copies $x_1$ to $y_1$. With this in mind, we are only interested in odd $p$, since they trivially generalize to even $p$.

In figures \ref{S3circuit}, \ref{S5circuit}, and \ref{S7circuit} below, we construct the circuits for the first few odd periods $S_3$, $S_5$ and $S_7$ respectively. The truth table for these circuits are included in tables \ref{S3table}, \ref{S5table} and \ref{S7table}. The circuits are constructed by inspection, and trial and error, making use of some general patterns and principles. Each circuit can be seen as two processes. The first is copying a linear combination of the input qubits to each output qubit via CNOT gates. This process must be used to create linearly independent combinations of the input bits in the output bits, which serve as the canvas on which the second process will operate. The linear independence insures the condition of injectivity within a single period is satisfied. 

The second process is the application of cascades of Toffoli gates to modify the results of the first process by flipping some entries in the truth table. The first Toffoli gate of the cascade uses two suitable control bits. The Toffoli gate at each subsequent level of the cascade uses as one of its control bits the target bit of the previous Toffoli gate. It is this Toffoli cascade process where most of the creativity lies, since it is what actually creates the desired periodicity. The first process alone cannot create odd periodicity. Note that the two processes may be made to occur in tandem. Also note that all output qubits $y_i$ are initialized to the state $\ket{0}$.

If one thinks of the truth tables below, the first Toffoli gate applied modifies a number of entries equal to a quarter of the total length of the column (i.e. $2^{n-2}$). Each consecutive Toffoli gate in the cascade flips half the number of entries of its target qubit as in the previous level of the cascade. For example, if $n=3$ (as is the case for $S_5$ and $S_7$), then the first Toffoli gate will flip $2^{3-2} = 2$ entries, the second gate in the cascade will flip 1 entry. If $n=4$, then the first Toffoli gate will flip $2^{4-2} = 4$ entries, the second gate in the cascade will flip 2 entries, and the third gate will flip 1 entry. For $n$ input qubits, a Toffoli cascade will have $n-1$ gates, with the final gate flipping only 1 entry in the truth table.

Where possible, the output of a Toffoli cascade is copied over to other qubits via CNOT gates, to avoid need for an identical cascade. This copying may take place in the middle of, or at the end of the cascade. Circuits for higher $p$ values may in addition have separate Toffoli gates that continue from where a copy of the cascade was made but in a ``different direction", or are not part of a cascade at all. Examples of this are provided in the appendix. 

To discuss the synthesis in more detail, we consider $S_3$, which implies $n=2$ bits each for the input and output registers. We see in figure \ref{S3circuit} that it requires 1 Toffoli gate, and 3 CNOT gates. The process is as follows, we set $y_1 = x_1 \oplus x_2$, where $\oplus$ denotes addition modulo 2 (i.e. the XOR operation). Then we use a Toffoli gate with $y_2$ as its target, which we then copy onto $y_1$. We use standard notation for CNOT and Toffoli gates, with a black circle indicating the control qubit, and a large circle with a plus sign inside indicates the target qubit. The small white circles we see in some circuits (such as fig. \ref{S5circuit}) are inverted control qubits, in the sense that the target bit is modified if the inverted control bit has value $0$, and is unchanged if it has value $1$.

The underlined entries in the truth table shown in table \ref{S3table} denote the entries flipped by the action of a Toffoli gate, whether directly or indirectly (where the result of the Toffoli is copied by a CNOT to another bit). Each underline is an entry flip, so an even number of underlines leaves the entry unchanged. The single horizontal line amid the truth table demarcates where the first period ends and the second period begins. The values of the output bits under this line must repeat the values at the top of the table.

%% S_3
\begin{figure}
\begin{floatrow}
\ffigbox{%
\centerline{
\Qcircuit @C=1em @R=.7em { 
\lstick{x_2} 		& \ctrl{3}	& \qw 		& \ctrl{2}  	&\qw		&\qw		& \rstick{x_2}    \\
\lstick{x_1}	 	& \qw 		& \ctrl{2}	& \qw  	&\qw		&\qw		& \rstick{x_1}    \\
\lstick{\ket{0}} 	& \qwc		& \qwc 		& \targc  	&\ctrlc{1}	&\qwc	& \rstick{y_2}    \\
\lstick{\ket{0}}	& \targc 	& \targc 	& \ctrlc{-1}  	&\targc	&\qwc		& \rstick{y_1}   \\ 
}}
}{%
  \fcaption{$S_{3}$ quantum circuit.}\label{S3circuit}%
}
\capbtabbox{%
\begin{tabular}{ l l | r r }  
$x_2$ & $x_1$ & $y_2$ & $y_1$ \\ 
  \hline \hline
  0 & 0 & 0 & 0 \\
  0 & 1 & 0 & 1 \\
  1 & 0 & \underline{1} & \underline{0} \\ \hline
  1 & 1 & 0 & 0 
\end{tabular}
}{%
  \tcaption{$S_{3}$ truth table.}\label{S3table}%
}
\end{floatrow}
\end{figure}
%% end S_3

Similarly, the circuit and truth table for $S_5$ are shown in fig. \ref{S5circuit} and table \ref{S5table} respectively. The circuit requires 2 Toffoli gates, and 3 CNOT gates. It involves setting  $y_1 = x_1 \oplus x_3$ and $y_2 = x_2$, then using a cascade of two Toffoli gates to flip $(2+1 =)$ 3 entries in the truth table. The modified entries are once again marked with an underline. Note that the only nonzero entry in the $y_3$ column of the truth table was not created by using CNOT additions from the $x_i$, as $y_3$ is not the target of any CNOT gates. Rather the value of $y_3$ was 'synthesized' through a cascade of Toffoli gates. This same idea emerges in many circuits as can be seen in the appendix.

The circuit and truth table for $S_7$ are shown in fig. \ref{S7circuit} and table \ref{S7table} respectively. The circuit requires 2 Toffoli gates, and 4 CNOT gates. Here we set $y_1 = x_1$ and $y_2 = x_2$, then we use a cascade of two Toffoli gates to flip 3 entries in the truth table, followed by a CNOT which duplicates one of these modified entries to another output bit, bringing the total number of flipped entries to 4, which are marked with an underline in the table. Finally, $x_3$ is added to $y_3$. Note that this final copying step was not done earlier in the process to facilitate the copying of the result of the second Toffoli gate from $y_3$ to another bit. 

%% S_5 and S_7
\begin{figure}
\begin{floatrow}
\ffigbox{%
\scalebox{1}{\centerline{
\Qcircuit @C=1em @R=.7em { 
\lstick{x_3} 		&\qw		& \qw 		& \ctrl{5}  	&\ctrl{4}	&\qw		&\qw		& \rstick{x_3}    \\
\lstick{x_2} 		&\qw		& \ctrl{3} 	& \qw  	&\qw		&\ctrlo{2}	&\qw		& \rstick{x_2}    \\
\lstick{x_1}	 	& \ctrl{3} 	& \qw		& \qw  	&\qw		&\qw		&\qw		& \rstick{x_1}    \\
\lstick{\ket{0}} 	& \qwc		& \qwc 		& \qwc  	&\qwc		&\targc		&\qwc		& \rstick{y_3}    \\
\lstick{\ket{0}} 	& \qwc		& \targc 	& \qwc  	&\targc		&\ctrlc{-1}	&\qwc		& \rstick{y_2}    \\
\lstick{\ket{0}}	& \targc 	& \qwc 		& \targc  	&\ctrlc{-1}	&\qwc		&\qwc		& \rstick{y_1}    
}}}
}{%
  \fcaption{$S_{5}$ quantum circuit.}\label{S5circuit}%
}
\ffigbox{%
\scalebox{1}{\centerline{
\Qcircuit @C=1em @R=.7em { 
\lstick{x_3} 		&\qw		& \ctrl{1}	& \qw  	&\qw		&\qw		&\ctrl{3}	&\qw		& \rstick{x_3}    \\
\lstick{x_2} 		&\qw		& \ctrl{4} 	& \qw  	&\qw		&\ctrl{3}	&\qw		&\qw		& \rstick{x_2}    \\
\lstick{x_1}	 	& \ctrl{3} 	& \qw		& \ctrl{2}  	&\qw		&\qw		&\qw		&\qw		& \rstick{x_1}    \\
\lstick{\ket{0}} 	& \qwc	& \qwc 	& \qwc  	&\targc	&\qwc		&\targc	&\qwc		& \rstick{y_3}    \\
\lstick{\ket{0}} 	& \qwc	& \qwc 	& \targc  	&\ctrlc{-1}	&\targc	&\qwc		&\qwc		& \rstick{y_2}    \\
\lstick{\ket{0}}	& \targc 	& \targc	& \ctrloc{-1} 	&\qwc		&\qwc		&\qwc		&\qwc		& \rstick{y_1}          
}}}
}{%
  \fcaption{$S_{7}$ quantum circuit.}\label{S7circuit}%
}
\end{floatrow}
% end circuits
%
\vspace{20pt}
%
% start tables
\begin{floatrow}
\capbtabbox{%
\scalebox{1}{
\begin{tabular}{ l l l | r r r }  
$x_3$ & $x_2$ & $x_1$ & $y_3$ & $y_2$ & $y_1$ \\ 
  \hline \hline
   0 & 0 & 0 & 0 & 0 & 0 \\
   0 & 0 & 1 & 0 & 0 & 1 \\
   0 & 1 & 0 & 0 & 1 & 0 \\
   0 & 1 & 1 & 0 & 1 & 1 \\
   1 & 0 & 0 & \underline{1} & \underline{1} & 1 \\ \hline
   1 & 0 & 1 & 0 & 0 & 0 \\
   1 & 1 & 0 & 0 & \underline{0} & 1 \\
   1 & 1 & 1 & 0 & 1 & 0 
\end{tabular}
}
}{%
  \tcaption{$S_{5}$ truth table.}\label{S5table}%
}
\capbtabbox{%
\scalebox{1}{
\begin{tabular}{ l l l | r r r }  
$x_3$ & $x_2$ & $x_1$ & $y_3$ & $y_2$ & $y_1$ \\ 
  \hline \hline
   0 & 0 & 0 & 0 & 0 & 0 \\
   0 & 0 & 1 & 0 & 0 & 1 \\
   0 & 1 & 0 & 0 & 1 & 0 \\
   0 & 1 & 1 & 0 & 1 & 1 \\
   1 & 0 & 0 & 1 & 0 & 0 \\ 
   1 & 0 & 1 & 1 & 0 & 1 \\
   1 & 1 & 0 & 1 & 1 & \underline{1} \\ \hline
   1 & 1 & 1 & \underline{0} & \underline{0} & \underline{0} \\
\end{tabular}
}
}{%
  \tcaption{$S_{7}$ truth table.}\label{S7table}%
}
\end{floatrow}
\end{figure}
% end S_5 and S_7

The circuit $S_9$ requires 3 Toffoli gates and 4 CNOT gates to construct, as shown in fig. \ref{S9circuit}. In a process very similar to the $S_5$ circuit, we set $y_1 = x_1 \oplus x_4, y_2 = x_2$, and $y_3 = x_3$. Then a cascade of three Toffoli gates is used to flip $(4+2+1$=) 7 entries in the truth table shown in table \ref{S9table}. Once again, the value of $y_4$ is synthesized solely using Toffoli gates, with no CNOT gates acting on that bit.

The circuit $S_{11}$ requires 4 Toffoli gates and 5 CNOT gates to construct in fig. \ref{S11circuit}. Here the construction of the circuit is more complicated. We start by setting $y_1 = x_1 \oplus x_4$ and $y_2=x_2$. We then follow it by a cascade of three Toffoli gates which flips 7 entries in table \ref{S11table}. We then add $x_3 \oplus x_4$ to $y_3$, which again we chose to do after the cascade of Toffoli gates. Finally, we add a fourth Toffoli gate, independent of the initial Toffoli cascade, to synthesize the contents of the $y_4$ column in the truth table. Overall, nine entries in the truth table have been modified by Toffoli gates. An entry with a double underline means its bit value was flipped twice, and therefore it was unchanged.

In the appendix we include the circuits and truth tables for $S_p$, for all odd $p$ up to $p=31$, i.e. up to $p$ a 5-bit number. The process of circuit construction is the same as the ones above, and we include some explanations and insights into any techniques used. 

There are interesting patterns in circuits for $p$ where $p=2^k \pm 1$ for any integer $k$. If $p=2^k + 1$, then the circuit for $S_p$ follows the pattern in $S_5, S_9,$ and $S_{17}$ shown in figures \ref{S5circuit}, \ref{S9circuit} and \ref{S17circuit}. The circuit will have $k+1$ input bits and the same number of output bits. It will start by copying each input bit $x_i$ to the output bit $y_i$ for $i=1, ..., k$. Then a single CNOT gates adds $x_{k+1}$ to $y_1$. Finally, a Toffoli cascade of $k$ gates is implemented, with bits $y_2$, $y_3$ .. $y_{k+1}$ as the target as each step. In total for such a circuit $N_T = k$ and $N_{CN}=k+1$.

If $p=2^k - 1$, then the circuit for $S_p$ follows the pattern in $S_7, S_{15},$ and $S_{31}$ shown in figures \ref{S7circuit}, \ref{S15circuit}, and \ref{S31circuit}. The circuit will have $k$ input bits and the same number of output bits. It will start by copying each input bit $x_i$ to the output bit $y_i$ for $i=1, ..., k-2$. Then a Toffoli cascade of $k-1$ gates, with the first gate having $x_k$ and $x_{k-1}$ as the control bits, and $y_1$ the target. The final target of the Toffoli cascade will be $y_{k-1}$, which is then copied by a CNOT to $y_k$. Finally $x_k$ and $x_{k-1}$ are added to $y_k$ and $y_{k-1}$ respectively. For this circuit $N_T = k-1$ and $N_{CN}=k+1$.

The appendix discusses other interesting patterns and relationships between the circuits, which may be exploited in the future to help create an optimal circuit synthesis algorithm.

%% S_9 and S_11
\begin{figure}[H]
\begin{floatrow}
\ffigbox{%
\scalebox{1}{\centerline{
\Qcircuit @C=1em @R=.7em { 
\lstick{x_4} 		&\qw		& \qw 		& \qw  	&\ctrl{7}	&\ctrl{6}	&\qw		&\qw		&\qw		& \rstick{x_4}    \\
\lstick{x_3} 		&\qw		& \qw 		& \ctrl{4}  	&\qw		&\qw		&\qw		&\ctrlo{3}	&\qw		& \rstick{x_3}    \\
\lstick{x_2} 		&\qw		& \ctrl{4} 	& \qw  	&\qw		&\qw		&\ctrlo{3}	&\qw		&\qw		& \rstick{x_2}    \\
\lstick{x_1}	 	& \ctrl{4} 	& \qw		& \qw  	&\qw		&\qw		&\qw		&\qw		&\qw		& \rstick{x_1}    \\
\lstick{\ket{0}} 	& \qwc		& \qwc 		& \qwc  	&\qwc		&\qwc		&\qwc		&\targc		&\qwc		& \rstick{y_4}    \\
\lstick{\ket{0}} 	& \qwc		& \qwc 		& \targc  	&\qwc		&\qwc		&\targc		&\ctrlc{-1}	&\qwc		& \rstick{y_3}    \\
\lstick{\ket{0}} 	& \qwc		& \targc 	& \qwc  	&\qwc		&\targc		&\ctrlc{-1}	&\qwc		&\qwc		& \rstick{y_2}    \\
\lstick{\ket{0}}	& \targc 	& \qwc 		& \qwc  	&\targc		&\ctrlc{-1}	&\qwc		&\qwc		&\qwc		& \rstick{y_1}       
}}}
}{%
  \fcaption{$S_{9}$ quantum circuit.}\label{S9circuit}%
}
\ffigbox{%
\scalebox{1}{\centerline{
\Qcircuit @C=1em @R=.7em { 
\lstick{x_4} 		&\qw		& \ctrl{7}	& \qw  	&\ctrl{3}	&\qw		&\qw		&\qw		&\ctrl{5}	&\ctrl{1}	&\qw		& \rstick{x_4}    \\
\lstick{x_3} 		&\qw		& \qw 		& \qw  	&\qw		&\qw		&\ctrlo{3}	&\ctrl{4}	&\qw		&\ctrlo{3}	&\qw		& \rstick{x_3}    \\
\lstick{x_2} 		&\qw		& \qw	 	& \ctrl{4}  	&\qw		&\ctrl{3}	&\qw		&\qw		&\qw		&\qw		&\qw		& \rstick{x_2}    \\
\lstick{x_1}	 	& \ctrl{4} 	& \qw		& \qw  	&\ctrl{3}	&\qw		&\qw		&\qw		&\qw		&\qw		&\qw		& \rstick{x_1}    \\
\lstick{\ket{0}} 	& \qwc		& \qwc 		& \qwc  	&\qwc		&\qwc		&\targc		&\qwc		&\qwc		&\targc		&\qwc		& \rstick{y_4}    \\
\lstick{\ket{0}} 	& \qwc		& \qwc 		& \qwc  	&\qwc		&\targc		&\ctrlc{-1}	&\targc		&\targc		&\qwc		&\qwc		& \rstick{y_3}    \\
\lstick{\ket{0}} 	& \qwc		& \qwc 		& \targc  	&\targc		&\ctrloc{-1}	&\qwc		&\qwc		&\qwc		&\qwc		&\qwc		& \rstick{y_2}    \\
\lstick{\ket{0}}	& \targc 	& \targc	& \qwc  	&\qwc		&\qwc		&\qwc		&\qwc		&\qwc		&\qwc		&\qwc		& \rstick{y_1}        
}}}
}{%
  \fcaption{$S_{11}$ quantum circuit.}\label{S11circuit}%
}
\end{floatrow}
% end circuits
%
\vspace{20pt}
%
% start tables
\begin{floatrow}
\capbtabbox{%
\scalebox{1}{
\begin{tabular}{l l l l | r r r r }  
$x_4$ & $x_3$ & $x_2$ & $x_1$ & $y_4$ & $y_3$ & $y_2$ & $y_1$ \\ 
  \hline \hline
   0 & 0 & 0 & 0 & 0 & 0 & 0 & 0 \\
   0 & 0 & 0 & 1 & 0 & 0 & 0 & 1 \\
   0 & 0 & 1 & 0 & 0 & 0 & 1 & 0 \\
   0 & 0 & 1 & 1 & 0 & 0 & 1 & 1 \\
   0 & 1 & 0 & 0 & 0 & 1 & 0 & 0 \\
   0 & 1 & 0 & 1 & 0 & 1 & 0 & 1 \\
   0 & 1 & 1 & 0 & 0 & 1 & 1 & 0 \\
   0 & 1 & 1 & 1 & 0 & 1 & 1 & 1 \\
   1 & 0 & 0 & 0 & \underline{1} & \underline{1} & \underline{1} & 1 \\ \hline
   1 & 0 & 0 & 1 & 0 & 0 & 0 & 0 \\
   1 & 0 & 1 & 0 & 0 & 0 & \underline{0} & 1 \\
   1 & 0 & 1 & 1 & 0 & 0 & 1 & 0 \\
   1 & 1 & 0 & 0 & 0 & \underline{0} & \underline{1} & 1 \\
   1 & 1 & 0 & 1 & 0 & 1 & 0 & 0 \\
   1 & 1 & 1 & 0 & 0 & 1 & \underline{0} & 1 \\
   1 & 1 & 1 & 1 & 0 & 1 & 1 & 0 \\
\end{tabular}
}
}{%
  \tcaption{$S_{9}$ truth table.}\label{S9table}%
}
\capbtabbox{%
\scalebox{1}{
\begin{tabular}{l l l l | r r r r }  
$x_4$ & $x_3$ & $x_2$ & $x_1$ & $y_4$ & $y_3$ & $y_2$ & $y_1$ \\ 
  \hline \hline
   0 & 0 & 0 & 0 & 0 & 0 & 0 & 0 \\
   0 & 0 & 0 & 1 & 0 & 0 & 0 & 1 \\
   0 & 0 & 1 & 0 & 0 & 0 & 1 & 0 \\
   0 & 0 & 1 & 1 & 0 & 0 & 1 & 1 \\
   0 & 1 & 0 & 0 & 0 & 1 & 0 & 0 \\
   0 & 1 & 0 & 1 & 0 & 1 & 0 & 1 \\
   0 & 1 & 1 & 0 & 0 & 1 & 1 & 0 \\
   0 & 1 & 1 & 1 & 0 & 1 & 1 & 1 \\
   1 & 0 & 0 & 0 & \underline{1} & 1 & 0 & 1 \\ 
   1 & 0 & 0 & 1 & \underline{1} & 1 & \underline{1} & 0 \\ 
   1 & 0 & 1 & 0 & \underline{1} & 1 & 1 & 1 \\ \hline
   1 & 0 & 1 & 1 & \underline{\underline{0}} & \underline{0} & \underline{0} & 0 \\
   1 & 1 & 0 & 0 & 0 & 0 & 0 & 1 \\
   1 & 1 & 0 & 1 & 0 & 0 & \underline{1} & 0 \\
   1 & 1 & 1 & 0 & 0 & 0 & 1 & 1 \\
   1 & 1 & 1 & 1 & 0 & \underline{1} & \underline{0} & 0 \\
\end{tabular}
}
}{%
  \tcaption{$S_{11}$ truth table.}\label{S11table}%
}
\end{floatrow}
\end{figure}
% end S_9 and S_11
%

\section{Required Resources for Synthesis}\label{sec4reqres}

One can follow the process illustrated in the previous section and expounded upon in the appendix to synthesize these circuits for arbitrary odd numbers. We have continued this process for larger circuits for period $p$ up to 5 bits. The required resources for each period have been summarized in table \ref{Ngatestable}, which provides the number of Toffoli and CNOT gates needed to synthesize the circuit of the function $S_p$. Note that $[p]_2$ is the period $p$ expressed in base 2.
\begin{table}[h]
\centerline{\footnotesize\
\begin{tabular}{r | r | r || r| r | r }  
Period $p$ & $[p]_2$ & $n$ & $N_T$ & $N_{CN}$ & $Q$\\ 
  \hline \hline
   3 & 11  & 2 & 1 & 3 & 9 \\ \hline
   5 & 101 & 3 & 2 & 3 & 15 \\ 
   7 & 111 & 3 & 2 & 4 & 16 \\ \hline
   9 & 1001 & 4 & 3 & 4 & 22 \\
   11 & 1011 & 4 & 4 & 5 & 29 \\
   13 & 1101 & 4 & 3 & 6 & 24 \\
   15 & 1111 & 4 & 3 & 5 & 23 \\ \hline
   17 & 10001 & 5 & 4 & 5 & 29 \\
   19 & 10011 & 5 & 5 & 6 & 36 \\
   21 & 10101 & 5 & 5 & 6 & 36 \\
   23 & 10111 &  5 & 5 & 7 & 37 \\
   25 & 11001 & 5 & 4 & 8 & 32 \\
   27 & 11011 & 5 & 5 & 7 & 37 \\
   29 & 11101 & 5 & 4 & 7 & 31 \\
   31 & 11111 & 5 & 4 & 6 & 30 \\
  \end{tabular}}
\tcaption{For each period $p$, we write the period in base 2 ($[p]_2$) and provide its bit-length $n$. The informative columns are the number of Toffoli gates ($N_T$) and CNOT gates ($N_{CN}$) needed to synthesize the $S_p$ circuit. We also include the quantum cost $Q = N_{CN} + 6N_{T}$.}\label{Ngatestable}
\end{table}
Given the above information, one can conjecture the following result of this paper: 

\begin{myconj*}
To synthesize the circuit for a simple periodic function $S_p$, where $p$  is an $n$-bit number, one needs at most $n$ Toffoli gates.
\end{myconj*} 

More precisely, let $[c]_2$ be the binary string equal to $[p]_2$ with the last bit truncated (since it is always $1$, because $p$ is odd). Then, we conjecture that for a given $p$, if the respective $[c]_2$ contains the substring $01$, which we call type A, then exactly $n$ Toffoli gates are needed for a simple periodic function. If the substring $01$ does not occur in $[c]_2$, which we call type B, then exactly $n-1$ Toffoli gates are needed. 

For a given bit-length $n$, there are $2^{n-2}$ odd periods $p$ where $p$ is an $n$-bit number. Of these $2^{n-2}$ possible odd periods, $n-1$ will be of type B, and the rest of type A. As an example, for $p= 23$, we have $[p]_2 = 10111$, then $[c]_2 = 1011$, which does contain the substring $01$, i.e. is type A, therefore $N_T=n=5$ Toffoli gates are needed. In the case $p= 25$, then $[p]_2 = 11001$, and $[c]_2 = 1100$, which does not contain the substring $01$, therefore it is type B and $N_T=n-1=4$ Toffoli gates are needed. 

The requirement that a certain substring be present in the binary representation of the period may seem a strange condition. However, its predictive power in the above examples seems to stem from the binary structure and recurrence of powers of 2 in the truth tables, particularly in the input columns. An exact analysis of the conjecture will be addressed in a future work.

\section{Conclusion}
We have defined an interesting class of simple periodic functions, with the intention that studying them yield some insights into periodicity in quantum circuits. We have demonstrated a custom procedure for circuit synthesis of $S_p$, a simple periodic function with period $p$, using only CNOT and Toffoli gates. 

We have provided examples of these circuits for many values of $p$. The procedure is immediately scalable to exactly construct circuits for periods $p$ of special forms $p = 2^k \pm 1$. For other $p$ values, the circuit synthesis procedure can be scaled on an ad-hoc basis. By analyzing the required resources for the synthesized circuits, we conjecture that for $p$ an $n$-bit number, one needs at most $n$ Toffoli gates to construct $S_p$. These simple periodic circuits may serve as stepping stones for experimental procedures as technology improves.

This leaves us with many interesting questions to address in subsequent work, such as the need for a scalable procedure for more general forms of $p$, and a proof for the conjecture above. A proof that a simple periodic function $S_p$ is the `simplest' possible (i.e. truly minimize the Toffoli gate count) is also of interest. Moreover, the periodic properties of these functions and their behaviour under Fourier transforms should be analyzed. Finally, we may ask how would one generalize the problem at hand to more complicated periodic functions with more than just one complete period.

\appendix{ Higher Periodic Circuits}

In this appendix, we list the circuits and truth tables for simple periodic functions $S_p$, for odd $p$ in the range $13 \le p \le 31$.

The circuits in this appendix and their associated truth tables follow similar patterns to the ones in section \ref{sec3circuitsyn}. However, some comments on a few of them are in order. The case of $S_{19}$, whose circuit is in fig. \ref{S19circuit} and associated truth table in table \ref{S19table}, demonstrates another interesting technique in circuit construction. The circuit starts with a Toffoli cascade on bits $y_2$, $y_3$, and $y_4$ that flip 8, 6, and 2 entries in the truth table respectively. Then another Toffoli gate from outside the cascade acts on $y_4$, flipping 8 entries, 2 of which were already flipped, resulting in an effective flip of $(8-2=)$ 6 entries. Another Toffoli gate then acts within the initial cascade, with the bit $y_5$ as its target, and flipping half the number in the previous level of the cascade, i.e. half of 6, and so 3 entries are flipped.

A pure Toffoli cascade can only flip a number of enties that is a power of 2 at each step. The technique used in the $S_{19}$ circuit demonstrates that one can interrupt the cascade with an 'independent' Toffoli gate to alter the number of entries flipped. The $S_{21}$ circuit in fig. \ref{S21circuit} uses the same technique, with the difference that the Toffoli gate that intervenes in the cascade just adds to the number of flipped entries, so we have $(8+2=)$ 10 entries flipped in the $y_4$ column, and half that number, 5 entries flipped in the last level of the cascade in the $y_5$ column.

%% S_13 and S_15
\begin{figure}[H]
\begin{floatrow}
\ffigbox{%
\scalebox{0.85}{\centerline{
\Qcircuit @C=1em @R=.7em { 
\lstick{x_4} 		&\ctrl{1}	& \qw		& \qw  	&\qw		&\qw		&\qw		&\qw		&\ctrl{4}	&\qw		&\qw		& \rstick{x_4}    \\
\lstick{x_3} 		&\ctrl{4}	& \qw 		& \qw  	&\qw		&\qw		&\qw		&\ctrl{4}	&\qw		&\ctrl{3}	&\qw		& \rstick{x_3}    \\
\lstick{x_2} 		&\qw		& \qw	 	& \qw  	&\ctrl{4}	&\qw		&\ctrlo{3}	&\qw		&\qw		&\qw		&\qw		& \rstick{x_2}    \\
\lstick{x_1}	 	& \qw 		& \qw		& \ctrl{4}  	&\qw		&\ctrlo{3}	&\qw		&\qw		&\qw		&\qw		&\qw		& \rstick{x_1}    \\
\lstick{\ket{0}} 	& \qwc	& \qwc 	& \qwc  	&\qwc		&\qwc		&\qwc		&\qwc		&\targc	&\targc	&\qwc		& \rstick{y_4}    \\
\lstick{\ket{0}} 	& \targc	& \ctrl{2} 	& \qwc  	&\qwc		&\qwc		&\targc	&\targc	&\qwc		&\qwc		&\qwc		& \rstick{y_3}    \\
\lstick{\ket{0}} 	& \qwc	& \qwc 	& \qwc  	&\targc	&\targc	&\ctrlc{-1}	&\qwc		&\qwc		&\qwc		&\qwc		& \rstick{y_2}    \\
\lstick{\ket{0}}	& \qwc 	& \targc	& \targc  	&\qwc		&\ctrlc{-1}	&\qwc		&\qwc		&\qwc		&\qwc		&\qwc		& \rstick{y_1}     
}}}
}{%
  \fcaption{$S_{13}$ quantum circuit.}\label{S13circuit}%
}
\ffigbox{%
\scalebox{0.85}{\centerline{
\Qcircuit @C=1em @R=.7em { 
\lstick{x_4} 		&\qw		& \qw		& \ctrl{1}  	&\qw		&\qw		&\qw		&\qw		&\ctrl{4}	&\qw		& \rstick{x_4}    \\
\lstick{x_3} 		&\qw		& \qw 		& \ctrl{6}  	&\qw		&\qw		&\qw		&\ctrl{4}	&\qw		&\qw		& \rstick{x_3}    \\
\lstick{x_2} 		&\qw		& \ctrl{4} 	& \qw  	&\qw		&\ctrl{3}	&\qw		&\qw		&\qw		&\qw		& \rstick{x_2}    \\
\lstick{x_1}	 	& \ctrl{4} 	& \qw		& \qw  	&\ctrl{3}	&\qw		&\qw		&\qw		&\qw		&\qw		& \rstick{x_1}    \\
\lstick{\ket{0}} 	& \qwc	& \qwc 	& \qwc  	&\qwc		&\qwc		&\targc	&\qwc		&\targc	&\qwc		& \rstick{y_4}    \\
\lstick{\ket{0}} 	& \qwc	& \qwc 	& \qwc  	&\qwc		&\targc	&\ctrlc{-1}	&\targc	&\qwc		&\qwc		& \rstick{y_3}    \\
\lstick{\ket{0}} 	& \qwc	& \targc  	& \qwc 	&\targc	&\ctrloc{-1}	&\qwc		&\qwc		&\qwc		&\qwc		& \rstick{y_2}    \\
\lstick{\ket{0}}	& \targc 	& \qwc	& \targc  	&\ctrloc{-1}	&\qwc		&\qwc		&\qwc		&\qwc		&\qwc		& \rstick{y_1}          
}}}
}{%
  \fcaption{$S_{15}$ quantum circuit.}\label{S15circuit}%
}
\end{floatrow}
% end circuits
%
\vspace{20pt}
%
% start tables
\begin{floatrow}
\capbtabbox{%
\scalebox{0.85}{
\begin{tabular}{l l l l | r r r r }  
$x_4$ & $x_3$ & $x_2$ & $x_1$ & $y_4$ & $y_3$ & $y_2$ & $y_1$ \\ 
  \hline \hline
   0 & 0 & 0 & 0 & 0 & 0 & 0 & 0 \\
   0 & 0 & 0 & 1 & 0 & 0 & 0 & 1 \\
   0 & 0 & 1 & 0 & 0 & 0 & 1 & 0 \\
   0 & 0 & 1 & 1 & 0 & 0 & 1 & 1 \\
   0 & 1 & 0 & 0 & 1 & 1 & 0 & 0 \\
   0 & 1 & 0 & 1 & 1 & 1 & 0 & 1 \\
   0 & 1 & 1 & 0 & 1 & 1 & 1 & 0 \\
   0 & 1 & 1 & 1 & 1 & 1 & 1 & 1 \\
   1 & 0 & 0 & 0 & 1 & 0 & 0 & 0 \\ 
   1 & 0 & 0 & 1 & 1 & 0 & 0 & 1 \\ 
   1 & 0 & 1 & 0 & 1 & 0 & 1 & 0 \\ 
   1 & 0 & 1 & 1 & 1 & 0 & 1 & 1 \\
   1 & 1 & 0 & 0 & 0 & \underline{\underline{1}} & \underline{1} & \underline{1} \\ \hline
   1 & 1 & 0 & 1 & 0 & \underline{0} & 0 & \underline{0} \\
   1 & 1 & 1 & 0 & 0 & \underline{0} & \underline{0} & \underline{1} \\
   1 & 1 & 1 & 1 & 0 & \underline{0} & 1 & \underline{0} \\
\end{tabular}
}
}{%
  \tcaption{$S_{13}$ truth table.}\label{S13table}%
}
\capbtabbox{%
\scalebox{0.85}{
\begin{tabular}{l l l l | r r r r }  
$x_4$ & $x_3$ & $x_2$ & $x_1$ & $y_4$ & $y_3$ & $y_2$ & $y_1$ \\ 
  \hline \hline
   0 & 0 & 0 & 0 & 0 & 0 & 0 & 0 \\
   0 & 0 & 0 & 1 & 0 & 0 & 0 & 1 \\
   0 & 0 & 1 & 0 & 0 & 0 & 1 & 0 \\
   0 & 0 & 1 & 1 & 0 & 0 & 1 & 1 \\
   0 & 1 & 0 & 0 & 0 & 1 & 0 & 0 \\
   0 & 1 & 0 & 1 & 0 & 1 & 0 & 1 \\
   0 & 1 & 1 & 0 & 0 & 1 & 1 & 0 \\
   0 & 1 & 1 & 1 & 0 & 1 & 1 & 1 \\
   1 & 0 & 0 & 0 & 1 & 0 & 0 & 0 \\ 
   1 & 0 & 0 & 1 & 1 & 0 & 0 & 1 \\ 
   1 & 0 & 1 & 0 & 1 & 0 & 1 & 0 \\ 
   1 & 0 & 1 & 1 & 1 & 0 & 1 & 1 \\
   1 & 1 & 0 & 0 & 1 & 1 & 0 & \underline{1} \\ 
   1 & 1 & 0 & 1 & 1 & 1 & \underline{1} & \underline{0} \\
   1 & 1 & 1 & 0 & 1 & 1 & 1 & \underline{1} \\ \hline
   1 & 1 & 1 & 1 & \underline{0} & \underline{0} & \underline{0} & \underline{0} \\
\end{tabular}
}
}{%
  \tcaption{$S_{15}$ truth table.}\label{S15table}%
}
\end{floatrow}
\end{figure}
% end S_13 and S_15

Interestingly, the circuits for $S_{19}$ and $S_{21}$ in figs. \ref{S19circuit} and \ref{S21circuit} respectively are almost identical, and only differ in the control value of some Toffoli gates (some black-filled circles denoting control bits are exchanged for white-filled ones, and vice versa).

%% S_17 and S_19
\begin{figure}[H]
\begin{floatrow}
\ffigbox{%
\scalebox{0.85}{\centerline{
\Qcircuit @C=1em @R=.7em { 
\lstick{x_5} 		&\qw		& \qw		& \qw  	&\qw		&\ctrl{9}	&\ctrl{8}	&\qw		&\qw		&\qw		&\qw		& \rstick{x_5}    \\
\lstick{x_4} 		&\qw		& \qw		& \qw  	&\ctrl{5}	&\qw		&\qw		&\qw		&\qw		&\ctrlo{4}	&\qw		& \rstick{x_4}    \\
\lstick{x_3} 		&\qw		& \qw 		& \ctrl{5}  	&\qw		&\qw		&\qw		&\qw		&\ctrlo{4}	&\qw		&\qw		& \rstick{x_3}    \\
\lstick{x_2} 		&\qw		& \ctrl{5} 	& \qw  	&\qw		&\qw		&\qw		&\ctrlo{4}	&\qw		&\qw		&\qw		& \rstick{x_2}    \\
\lstick{x_1}	 	& \ctrl{5} 	& \qw		& \qw  	&\qw		&\qw		&\qw		&\qw		&\qw		&\qw		&\qw		& \rstick{x_1}    \\
\lstick{\ket{0}} 	& \qwc	& \qwc 	& \qwc  	&\qwc		&\qwc		&\qwc		&\qwc		&\qwc		&\targc	&\qwc		& \rstick{y_5}    \\
\lstick{\ket{0}} 	& \qwc	& \qwc 	& \qwc  	&\targc	&\qwc		&\qwc		&\qwc		&\targc	&\ctrlc{-1}	&\qwc		& \rstick{y_4}    \\
\lstick{\ket{0}} 	& \qwc	& \qwc 	& \targc  	&\qwc		&\qwc		&\qwc		&\targc	&\ctrlc{-1}	&\qwc		&\qwc		& \rstick{y_3}    \\
\lstick{\ket{0}} 	& \qwc	& \targc 	& \qwc  	&\qwc		&\qwc		&\targc	&\ctrlc{-1}	&\qwc		&\qwc		&\qwc		& \rstick{y_2}    \\
\lstick{\ket{0}}	& \targc 	& \qwc	& \qwc  	&\qwc		&\targc	&\ctrlc{-1}	&\qwc		&\qwc		&\qwc		&\qwc		& \rstick{y_1}     
}}}
}{%
  \fcaption{$S_{17}$ quantum circuit.}\label{S17circuit}%
}
\ffigbox{%
\scalebox{0.85}{\centerline{
\Qcircuit @C=.8em @R=.7em { 
\lstick{x_5} 		&\ctrl{4}	& \qw		&\qw  		&\ctrl{2}	&\qw		&\qw		&\ctrl{9}	&\qw		&\qw		&\ctrl{7}	&\qw		&\qw		& \rstick{x_5}    \\
\lstick{x_4} 		&\qw		& \qw		&\qw  		&\qw		&\ctrlo{4}	&\qw		&\qw		&\qw		&\qw		&\qw		&\ctrl{5}	&\qw		& \rstick{x_4}    \\
\lstick{x_3} 		&\qw		& \qw 		&\ctrlo{4}  	&\ctrlo{4}	&\qw		&\qw		&\qw		&\qw		&\ctrl{5}	&\qw		&\qw		&\qw		& \rstick{x_3}    \\
\lstick{x_2} 		&\qw		& \ctrl{4} 	& \qw  	&\qw		&\qw		&\qw		&\qw		&\ctrl{5}	&\qw		&\qw		&\qw		&\qw		& \rstick{x_2}    \\
\lstick{x_1}	 	& \ctrl{4} 	& \qw		& \qw  	&\qw		&\qw		&\ctrl{5}	&\qw		&\qw		&\qw		&\qw		&\qw		&\qw		& \rstick{x_1}    \\
\lstick{\ket{0}} 	& \qwc	& \qwc 	& \qwc  	&\qwc		&\targc	&\qwc		&\qwc		&\qwc		&\qwc		&\qwc		&\qwc		&\qwc		& \rstick{y_5}    \\
\lstick{\ket{0}} 	& \qwc	& \qwc 	& \targc  	&\targc	&\ctrlc{-1}	&\qwc		&\qwc		&\qwc		&\qwc		&\qwc		&\targc	&\qwc		& \rstick{y_4}    \\
\lstick{\ket{0}} 	& \qwc	& \targc 	& \ctrlc{-1}  	&\qwc		&\qwc		&\qwc		&\qwc		&\qwc		&\targc	&\targc	&\qwc		&\qwc		& \rstick{y_3}    \\
\lstick{\ket{0}} 	& \targc	& \ctrlc{-1} 	& \qwc  	&\qwc		&\qwc		&\qwc		&\qwc		&\targc	&\qwc		&\qwc		&\qwc		&\qwc		& \rstick{y_2}    \\
\lstick{\ket{0}}	& \qwc 	& \qwc	& \qwc  	&\qwc		&\qwc		&\targc	&\targc	&\qwc		&\qwc		&\qwc		&\qwc		&\qwc		& \rstick{y_1}           
}}}
}{%
  \fcaption{$S_{19}$ quantum circuit.}\label{S19circuit}%
}
\end{floatrow}
% end circuits
%
\vspace{20pt}
%
% start tables
\begin{floatrow}
\capbtabbox{%
\scalebox{0.85}{
\begin{tabular}{l l l l l | r r r r r }  
$x_5$ & $x_4$ & $x_3$ & $x_2$ & $x_1$ & $y_5$ & $y_4$ & $y_3$ & $y_2$ & $y_1$ \\ 
  \hline \hline
   0 & 0 & 0 & 0 & 0 & 0 & 0 & 0 & 0 & 0 \\
   0 & 0 & 0 & 0 & 1 & 0 & 0 & 0 & 0 & 1 \\
   0 & 0 & 0 & 1 & 0 & 0 & 0 & 0 & 1 & 0 \\
   0 & 0 & 0 & 1 & 1 & 0 & 0 & 0 & 1 & 1 \\
   0 & 0 & 1 & 0 & 0 & 0 & 0 & 1 & 0 & 0 \\
   0 & 0 & 1 & 0 & 1 & 0 & 0 & 1 & 0 & 1 \\
   0 & 0 & 1 & 1 & 0 & 0 & 0 & 1 & 1 & 0 \\
   0 & 0 & 1 & 1 & 1 & 0 & 0 & 1 & 1 & 1 \\
   0 & 1 & 0 & 0 & 0 & 0 & 1 & 0 & 0 & 0 \\ 
   0 & 1 & 0 & 0 & 1 & 0 & 1 & 0 & 0 & 1 \\ 
   0 & 1 & 0 & 1 & 0 & 0 & 1 & 0 & 1 & 0 \\ 
   0 & 1 & 0 & 1 & 1 & 0 & 1 & 0 & 1 & 1 \\
   0 & 1 & 1 & 0 & 0 & 0 & 1 & 1 & 0 & 0 \\ 
   0 & 1 & 1 & 0 & 1 & 0 & 1 & 1 & 0 & 1 \\
   0 & 1 & 1 & 1 & 0 & 0 & 1 & 1 & 1 & 0 \\ 
   0 & 1 & 1 & 1 & 1 & 0 & 1 & 1 & 1 & 1 \\
   1 & 0 & 0 & 0 & 0 & \underline{1} & \underline{1} & \underline{1} & \underline{1} & 1 \\ \hline
   1 & 0 & 0 & 0 & 1 & 0 & 0 & 0 & 0 & 0 \\
   1 & 0 & 0 & 1 & 0 & 0 & 0 & 0 & \underline{0} & 1 \\ 
   1 & 0 & 0 & 1 & 1 & 0 & 0 & 0 & 1 & 0 \\
   1 & 0 & 1 & 0 & 0 & 0 & 0 & \underline{0} & \underline{1} & 1 \\
   1 & 0 & 1 & 0 & 1 & 0 & 0 & 1 & 0 & 0 \\
   1 & 0 & 1 & 1 & 0 & 0 & 0 & 1 & \underline{0} & 1 \\
   1 & 0 & 1 & 1 & 1 & 0 & 0 & 1 & 1 & 0 \\
   1 & 1 & 0 & 0 & 0 & 0 & \underline{0} & \underline{1} & \underline{1} & 1 \\ 
   1 & 1 & 0 & 0 & 1 & 0 & 1 & 0 & 0 & 0 \\ 
   1 & 1 & 0 & 1 & 0 & 0 & 1 & 0 & \underline{0} & 1 \\ 
   1 & 1 & 0 & 1 & 1 & 0 & 1 & 0 & 1 & 0 \\
   1 & 1 & 1 & 0 & 0 & 0 & 1 & \underline{0} & \underline{1} & 1 \\ 
   1 & 1 & 1 & 0 & 1 & 0 & 1 & 1 & 0 & 0 \\
   1 & 1 & 1 & 1 & 0 & 0 & 1 & 1 & \underline{0} & 1 \\ 
   1 & 1 & 1 & 1 & 1 & 0 & 1 & 1 & 1 & 0 \\
\end{tabular}
}
}{%
  \tcaption{$S_{17}$ truth table.}\label{S17table}%
}
\capbtabbox{%
\scalebox{0.85}{
\begin{tabular}{l l l l l | r r r r r }  
$x_5$ & $x_4$ & $x_3$ & $x_2$ & $x_1$ & $y_5$ & $y_4$ & $y_3$ & $y_2$ & $y_1$ \\ 
  \hline \hline
   0 & 0 & 0 & 0 & 0 & 0 & 0 & 0 & 0 & 0 \\
   0 & 0 & 0 & 0 & 1 & 0 & 0 & 0 & 0 & 1 \\
   0 & 0 & 0 & 1 & 0 & 0 & 0 & 0 & 1 & 0 \\
   0 & 0 & 0 & 1 & 1 & 0 & 0 & 0 & 1 & 1 \\
   0 & 0 & 1 & 0 & 0 & 0 & 0 & 1 & 0 & 0 \\
   0 & 0 & 1 & 0 & 1 & 0 & 0 & 1 & 0 & 1 \\
   0 & 0 & 1 & 1 & 0 & 0 & 0 & 1 & 1 & 0 \\
   0 & 0 & 1 & 1 & 1 & 0 & 0 & 1 & 1 & 1 \\
   0 & 1 & 0 & 0 & 0 & 0 & 1 & 0 & 0 & 0 \\ 
   0 & 1 & 0 & 0 & 1 & 0 & 1 & 0 & 0 & 1 \\ 
   0 & 1 & 0 & 1 & 0 & 0 & 1 & 0 & 1 & 0 \\ 
   0 & 1 & 0 & 1 & 1 & 0 & 1 & 0 & 1 & 1 \\
   0 & 1 & 1 & 0 & 0 & 0 & 1 & 1 & 0 & 0 \\ 
   0 & 1 & 1 & 0 & 1 & 0 & 1 & 1 & 0 & 1 \\
   0 & 1 & 1 & 1 & 0 & 0 & 1 & 1 & 1 & 0 \\ 
   0 & 1 & 1 & 1 & 1 & 0 & 1 & 1 & 1 & 1 \\
   1 & 0 & 0 & 0 & 0 & \underline{1} & \underline{1} & 1 & 0 & 1 \\ 
   1 & 0 & 0 & 0 & 1 & \underline{1} & \underline{1} & 1 & \underline{1} & 0 \\
   1 & 0 & 0 & 1 & 0 & \underline{1} & \underline{1} & 1 & 1 & 1 \\ \hline
   1 & 0 & 0 & 1 & 1 & \underline{\underline{0}} & \underline{\underline{0}} & \underline{0} & \underline{0} & 0 \\
   1 & 0 & 1 & 0 & 0 & 0 & 0 & 0 & 0 & 1 \\ 
   1 & 0 & 1 & 0 & 1 & 0 & 0 & 0 & \underline{1} & 0 \\
   1 & 0 & 1 & 1 & 0 & 0 & 0 & 0 & 1 & 1 \\
   1 & 0 & 1 & 1 & 1 & 0 & 0 & \underline{1} & \underline{0} & 0 \\
   1 & 1 & 0 & 0 & 0 & 0 & \underline{0} & 1 & 0 & 1 \\ 
   1 & 1 & 0 & 0 & 1 & 0 & \underline{0} & 1 & \underline{1} & 0 \\ 
   1 & 1 & 0 & 1 & 0 & 0 & \underline{0} & 1 & 1 & 1 \\ 
   1 & 1 & 0 & 1 & 1 & 0 & \underline{\underline{1}} & \underline{0} & \underline{0} & 0 \\
   1 & 1 & 1 & 0 & 0 & 0 & 1 & 0 & 0 & 1 \\ 
   1 & 1 & 1 & 0 & 1 & 0 & 1 & 0 & \underline{1} & 0 \\
   1 & 1 & 1 & 1 & 0 & 0 & 1 & 0 & 1 & 1 \\ 
   1 & 1 & 1 & 1 & 1 & 0 & 1 & \underline{1} & \underline{0} & 0 \\
\end{tabular}
}
}{%
  \tcaption{$S_{19}$ truth table.}\label{S19table}%
}
\end{floatrow}
\end{figure}
% end S_17 and S_19

%% S_21 and S_23
\begin{figure}[H]
\begin{floatrow}
\ffigbox{%
\scalebox{0.85}{\centerline{
\Qcircuit @C=.8em @R=.7em { 
\lstick{x_5} 		&\ctrl{4}	& \qw		&\qw  		&\ctrl{2}	&\qw		&\qw		&\ctrl{9}	&\qw		&\qw		&\ctrl{7}	&\qw		&\qw		& \rstick{x_5}    \\
\lstick{x_4} 		&\qw		& \qw		&\qw  		&\qw		&\ctrlo{4}	&\qw		&\qw		&\qw		&\qw		&\qw		&\ctrl{5}	&\qw		& \rstick{x_4}    \\
\lstick{x_3} 		&\qw		& \qw 		&\ctrl{4}  	&\ctrlo{4}	&\qw		&\qw		&\qw		&\qw		&\ctrl{5}	&\qw		&\qw		&\qw		& \rstick{x_3}    \\
\lstick{x_2} 		&\qw		& \ctrlo{4} 	& \qw  	&\qw		&\qw		&\qw		&\qw		&\ctrl{5}	&\qw		&\qw		&\qw		&\qw		& \rstick{x_2}    \\
\lstick{x_1}	 	& \ctrlo{4} 	& \qw		& \qw  	&\qw		&\qw		&\ctrl{5}	&\qw		&\qw		&\qw		&\qw		&\qw		&\qw		& \rstick{x_1}    \\
\lstick{\ket{0}} 	& \qwc	& \qwc 	& \qwc  	&\qwc		&\targc	&\qwc		&\qwc		&\qwc		&\qwc		&\qwc		&\qwc		&\qwc		& \rstick{y_5}    \\
\lstick{\ket{0}} 	& \qwc	& \qwc 	& \targc  	&\targc	&\ctrlc{-1}	&\qwc		&\qwc		&\qwc		&\qwc		&\qwc		&\targc	&\qwc		& \rstick{y_4}    \\
\lstick{\ket{0}} 	& \qwc	& \targc 	& \ctrlc{-1}  	&\qwc		&\qwc		&\qwc		&\qwc		&\qwc		&\targc	&\targc	&\qwc		&\qwc		& \rstick{y_3}    \\
\lstick{\ket{0}} 	& \targc	& \ctrlc{-1} 	& \qwc  	&\qwc		&\qwc		&\qwc		&\qwc		&\targc	&\qwc		&\qwc		&\qwc		&\qwc		& \rstick{y_2}    \\
\lstick{\ket{0}}	& \qwc 	& \qwc	& \qwc  	&\qwc		&\qwc		&\targc	&\targc	&\qwc		&\qwc		&\qwc		&\qwc		&\qwc		& \rstick{y_1}          
}}}
}{%
  \fcaption{$S_{21}$ quantum circuit.}\label{S21circuit}%
}
\ffigbox{%
\scalebox{0.85}{\centerline{
\Qcircuit @C=.7em @R=.7em { 
\lstick{x_5} 		&\ctrl{4}	& \qw		&\qw  		&\qw		&\ctrl{1}	&\qw		&\ctrl{9}	&\qw		&\qw		&\qw		&\qw		&\ctrl{5}	&\qw		& \rstick{x_5}    \\
\lstick{x_4} 		&\qw		& \qw		&\qw  		&\ctrl{4}	&\ctrl{5}	&\qw		&\qw		&\qw		&\qw		&\ctrl{5}	&\ctrl{4}	&\qw		&\qw		& \rstick{x_4}    \\
\lstick{x_3} 		&\qw		& \qw 		&\ctrl{3}  	&\qw		&\qw		&\qw		&\qw		&\qw		&\ctrl{5}	&\qw		&\qw		&\qw		&\qw		& \rstick{x_3}    \\
\lstick{x_2} 		&\qw		& \ctrl{4} 	& \qw  	&\qw		&\qw		&\qw		&\qw		&\ctrl{5}	&\qw		&\qw		&\qw		&\qw		&\qw		& \rstick{x_2}    \\
\lstick{x_1}	 	& \ctrl{4} 	& \qw		& \qw  	&\qw		&\qw		&\ctrl{5}	&\qw		&\qw		&\qw		&\qw		&\qw		&\qw		&\qw		& \rstick{x_1}    \\
\lstick{\ket{0}} 	& \qwc	& \qwc 	& \targc  	&\ctrlc{1}	&\qwc		&\qwc		&\qwc		&\qwc		&\qwc		&\qwc		&\targc	&\targc	&\qwc		& \rstick{y_5}    \\
\lstick{\ket{0}} 	& \qwc	& \qwc 	& \qwc  	&\targc	&\targc	&\qwc		&\qwc		&\qwc		&\qwc		&\targc	&\qwc		&\qwc		&\qwc		& \rstick{y_4}    \\
\lstick{\ket{0}} 	& \qwc	& \targc 	& \ctrlc{-2}  	&\qwc		&\qwc		&\qwc		&\qwc		&\qwc		&\targc	&\qwc		&\qwc		&\qwc		&\qwc		& \rstick{y_3}    \\
\lstick{\ket{0}} 	& \targc	& \ctrlc{-1} 	& \qwc  	&\qwc		&\qwc		&\qwc		&\qwc		&\targc	&\qwc		&\qwc		&\qwc		&\qwc		&\qwc		& \rstick{y_2}    \\
\lstick{\ket{0}}	& \qwc 	& \qwc	& \qwc  	&\qwc		&\qwc		&\targc	&\targc	&\qwc		&\qwc		&\qwc		&\qwc		&\qwc		&\qwc		& \rstick{y_1}       
}}}
}{%
  \fcaption{$S_{23}$ quantum circuit.}\label{S23circuit}%
}
\end{floatrow}
% end circuits
%
\vspace{20pt}
%
% start tables
\begin{floatrow}
\capbtabbox{%
\scalebox{0.85}{
\begin{tabular}{l l l l l | r r r r r }  
$x_5$ & $x_4$ & $x_3$ & $x_2$ & $x_1$ & $y_5$ & $y_4$ & $y_3$ & $y_2$ & $y_1$ \\ 
  \hline \hline
   0 & 0 & 0 & 0 & 0 & 0 & 0 & 0 & 0 & 0 \\
   0 & 0 & 0 & 0 & 1 & 0 & 0 & 0 & 0 & 1 \\
   0 & 0 & 0 & 1 & 0 & 0 & 0 & 0 & 1 & 0 \\
   0 & 0 & 0 & 1 & 1 & 0 & 0 & 0 & 1 & 1 \\
   0 & 0 & 1 & 0 & 0 & 0 & 0 & 1 & 0 & 0 \\
   0 & 0 & 1 & 0 & 1 & 0 & 0 & 1 & 0 & 1 \\
   0 & 0 & 1 & 1 & 0 & 0 & 0 & 1 & 1 & 0 \\
   0 & 0 & 1 & 1 & 1 & 0 & 0 & 1 & 1 & 1 \\
   0 & 1 & 0 & 0 & 0 & 0 & 1 & 0 & 0 & 0 \\ 
   0 & 1 & 0 & 0 & 1 & 0 & 1 & 0 & 0 & 1 \\ 
   0 & 1 & 0 & 1 & 0 & 0 & 1 & 0 & 1 & 0 \\ 
   0 & 1 & 0 & 1 & 1 & 0 & 1 & 0 & 1 & 1 \\
   0 & 1 & 1 & 0 & 0 & 0 & 1 & 1 & 0 & 0 \\ 
   0 & 1 & 1 & 0 & 1 & 0 & 1 & 1 & 0 & 1 \\
   0 & 1 & 1 & 1 & 0 & 0 & 1 & 1 & 1 & 0 \\ 
   0 & 1 & 1 & 1 & 1 & 0 & 1 & 1 & 1 & 1 \\
   1 & 0 & 0 & 0 & 0 & \underline{1} & \underline{1} & \underline{0} & \underline{1} & 1 \\ 
   1 & 0 & 0 & 0 & 1 & \underline{1} & \underline{1} & 1 & 0 & 0 \\
   1 & 0 & 0 & 1 & 0 & \underline{1} & \underline{1} & 1 & \underline{0} & 1 \\ 
   1 & 0 & 0 & 1 & 1 & \underline{1} & \underline{1} & 1 & 1 & 0 \\
   1 & 0 & 1 & 0 & 0 & \underline{1} & \underline{1} & \underline{1} & \underline{1} & 1 \\ \hline
   1 & 0 & 1 & 0 & 1 & 0 & 0 & 0 & 0 & 0 \\
   1 & 0 & 1 & 1 & 0 & 0 & 0 & 0 & \underline{0} & 1 \\
   1 & 0 & 1 & 1 & 1 & 0 & 0 & 0 & 1 & 0 \\
   1 & 1 & 0 & 0 & 0 & 0 & \underline{0} & \underline{0} & \underline{1} & 1 \\ 
   1 & 1 & 0 & 0 & 1 & 0 & \underline{0} & 1 & 0 & 0 \\ 
   1 & 1 & 0 & 1 & 0 & 0 & \underline{0} & 1 & \underline{0} & 1 \\ 
   1 & 1 & 0 & 1 & 1 & 0 & \underline{0} & 1 & 1 & 0 \\
   1 & 1 & 1 & 0 & 0 & 0 & \underline{0} & \underline{1} & \underline{1} & 1 \\ 
   1 & 1 & 1 & 0 & 1 & 0 & 1 & 0 & 0 & 0 \\
   1 & 1 & 1 & 1 & 0 & 0 & 1 & 0 & \underline{0} & 1 \\ 
   1 & 1 & 1 & 1 & 1 & 0 & 1 & 0 & 1 & 0 \\
\end{tabular}
}
}{%
  \tcaption{$S_{21}$ truth table.}\label{S21table}%
}
\capbtabbox{%
\scalebox{0.85}{
\begin{tabular}{l l l l l | r r r r r }  
$x_5$ & $x_4$ & $x_3$ & $x_2$ & $x_1$ & $y_5$ & $y_4$ & $y_3$ & $y_2$ & $y_1$ \\ 
  \hline \hline
   0 & 0 & 0 & 0 & 0 & 0 & 0 & 0 & 0 & 0 \\
   0 & 0 & 0 & 0 & 1 & 0 & 0 & 0 & 0 & 1 \\
   0 & 0 & 0 & 1 & 0 & 0 & 0 & 0 & 1 & 0 \\
   0 & 0 & 0 & 1 & 1 & 0 & 0 & 0 & 1 & 1 \\
   0 & 0 & 1 & 0 & 0 & 0 & 0 & 1 & 0 & 0 \\
   0 & 0 & 1 & 0 & 1 & 0 & 0 & 1 & 0 & 1 \\
   0 & 0 & 1 & 1 & 0 & 0 & 0 & 1 & 1 & 0 \\
   0 & 0 & 1 & 1 & 1 & 0 & 0 & 1 & 1 & 1 \\
   0 & 1 & 0 & 0 & 0 & 1 & 1 & 0 & 0 & 0 \\ 
   0 & 1 & 0 & 0 & 1 & 1 & 1 & 0 & 0 & 1 \\ 
   0 & 1 & 0 & 1 & 0 & 1 & 1 & 0 & 1 & 0 \\ 
   0 & 1 & 0 & 1 & 1 & 1 & 1 & 0 & 1 & 1 \\
   0 & 1 & 1 & 0 & 0 & 1 & 1 & 1 & 0 & 0 \\ 
   0 & 1 & 1 & 0 & 1 & 1 & 1 & 1 & 0 & 1 \\
   0 & 1 & 1 & 1 & 0 & 1 & 1 & 1 & 1 & 0 \\ 
   0 & 1 & 1 & 1 & 1 & 1 & 1 & 1 & 1 & 1 \\
   1 & 0 & 0 & 0 & 0 & 1 & 0 & 0 & 0 & 1 \\ 
   1 & 0 & 0 & 0 & 1 & 1 & 0 & 0 & \underline{1} & 0 \\
   1 & 0 & 0 & 1 & 0 & 1 & 0 & 0 & 1 & 1 \\ 
   1 & 0 & 0 & 1 & 1 & 1 & 0 & \underline{1} & \underline{0} & 0 \\
   1 & 0 & 1 & 0 & 0 & 1 & 0 & 1 & 0 & 1 \\ 
   1 & 0 & 1 & 0 & 1 & 1 & 0 & 1 & \underline{1} & 0 \\
   1 & 0 & 1 & 1 & 0 & 1 & 0 & 1 & 1 & 1 \\ \hline
   1 & 0 & 1 & 1 & 1 & \underline{0} & 0 & \underline{0} & \underline{0} & 0 \\
   1 & 1 & 0 & 0 & 0 & 0 & \underline{0} & 0 & 0 & 1 \\ 
   1 & 1 & 0 & 0 & 1 & 0 & \underline{0} & 0 & \underline{1} & 0 \\ 
   1 & 1 & 0 & 1 & 0 & 0 & \underline{0} & 0 & 1 & 1 \\ 
   1 & 1 & 0 & 1 & 1 & 0 & \underline{0} & \underline{1} & \underline{0} & 0 \\
   1 & 1 & 1 & 0 & 0 & 0 & \underline{0} & 1 & 0 & 1 \\ 
   1 & 1 & 1 & 0 & 1 & 0 & \underline{0} & 1 & \underline{1} & 0 \\
   1 & 1 & 1 & 1 & 0 & 0 & \underline{0} & 1 & 1 & 1 \\ 
   1 & 1 & 1 & 1 & 1 & \underline{1} & \underline{\underline{1}} & \underline{0} & \underline{0} & 0 \\
\end{tabular}
}
}{%
  \tcaption{$S_{23}$ truth table.}\label{S23table}%
}
\end{floatrow}
\end{figure}
% end S_21 and S_23

%% S_25 and S_27
\begin{figure}[H]
\begin{floatrow}
\ffigbox{%
\scalebox{0.85}{\centerline{
\Qcircuit @C=.7em @R=.7em { 
\lstick{x_5} 		&\ctrl{1}	& \qw		&\qw  		&\qw		&\qw		&\qw		&\qw		&\qw		&\qw		&\qw		&\qw		&\ctrl{5}	&\qw		& \rstick{x_5}    \\
\lstick{x_4} 		&\ctrl{8}	& \qw		&\qw  		&\qw		&\qw		&\qw		&\qw		&\qw		&\qw		&\ctrl{5}	&\ctrl{4}	&\qw		&\qw		& \rstick{x_4}    \\
\lstick{x_3} 		&\qw		& \qw 		&\qw  		&\ctrl{4}	&\qw		&\qw		&\qw		&\qw		&\ctrl{5}	&\qw		&\qw		&\qw		&\qw		& \rstick{x_3}    \\
\lstick{x_2} 		&\qw		& \qw 		& \ctrlo{3}  	&\qw		&\qw		&\qw		&\qw		&\ctrl{5}	&\qw		&\qw		&\qw		&\qw		&\qw		& \rstick{x_2}    \\
\lstick{x_1}	 	& \qw 		& \ctrlo{4}	& \qw  	&\qw		&\qw		&\qw		&\ctrl{5}	&\qw		&\qw		&\qw		&\qw		&\qw		&\qw		& \rstick{x_1}    \\
\lstick{\ket{0}} 	& \qwc	& \qwc 	& \qwc  	&\qwc		&\qwc		&\qwc		&\qwc		&\qwc		&\qwc		&\qwc		&\targc	&\targc	&\qwc		& \rstick{y_5}    \\
\lstick{\ket{0}} 	& \qwc	& \qwc 	& \targc  	&\ctrlc{1}	&\targc	&\targc	&\qwc		&\qwc		&\qwc		&\targc	&\qwc		&\qwc		&\qwc		& \rstick{y_4}    \\
\lstick{\ket{0}} 	& \qwc	& \qwc 	& \qwc  	&\targc	&\ctrlc{-1}	&\qwc		&\qwc		&\qwc		&\targc	&\qwc		&\qwc		&\qwc		&\qwc		& \rstick{y_3}    \\
\lstick{\ket{0}} 	& \qwc	& \targc	& \ctrlc{-2}  	&\qwc		&\qwc		&\qwc		&\qwc		&\targc	&\qwc		&\qwc		&\qwc		&\qwc		&\qwc		& \rstick{y_2}    \\
\lstick{\ket{0}}	& \targc 	& \ctrlc{-1}	& \qwc  	&\qwc		&\qwc		&\ctrlc{-3}	&\targc	&\qwc		&\qwc		&\qwc		&\qwc		&\qwc		&\qwc		& \rstick{y_1}            
}}}
}{%
  \fcaption{$S_{25}$ quantum circuit.}\label{S25circuit}%
}
\ffigbox{%
\scalebox{0.85}{\centerline{
\Qcircuit @C=.7em @R=.7em { 
\lstick{x_5} 		&\ctrl{1}	& \qw		&\qw  		&\qw		&\qw		&\qw		&\qw		&\qw		&\qw		&\qw		&\qw		&\ctrl{5}	&\qw		& \rstick{x_5}    \\
\lstick{x_4} 		&\ctrl{8}	& \qw		&\qw  		&\qw		&\qw		&\qw		&\qw		&\qw		&\ctrl{5}	&\qw		&\ctrl{4}	&\qw		&\qw		& \rstick{x_4}    \\
\lstick{x_3} 		&\qw		& \qw 		&\ctrl{5}	&\qw		&\ctrl{4}	&\qw		&\qw		&\ctrl{5}	&\qw		&\ctrl{4}	&\qw		&\qw		&\qw		& \rstick{x_3}    \\
\lstick{x_2} 		&\qw		& \qw 		& \qw  	&\ctrl{3}	&\qw		&\qw		&\ctrl{5}	&\qw		&\qw		&\qw		&\qw		&\qw		&\qw		& \rstick{x_2}    \\
\lstick{x_1}	 	& \qw 		& \ctrl{4}	& \qw  	&\qw		&\qw		&\ctrl{5}	&\qw		&\qw		&\qw		&\qw		&\qw		&\qw		&\qw		& \rstick{x_1}    \\
\lstick{\ket{0}} 	& \qwc	& \qwc 	& \qwc  	&\qwc		&\qwc		&\qwc		&\qwc		&\qwc		&\qwc		&\qwc		&\targc	&\targc	&\qwc		& \rstick{y_5}    \\
\lstick{\ket{0}} 	& \qwc	& \qwc 	& \qwc  	&\targc	&\ctrlc{1}	&\qwc		&\qwc		&\qwc		&\targc	&\targc	&\qwc		&\qwc		&\qwc		& \rstick{y_4}    \\
\lstick{\ket{0}} 	& \qwc	& \qwc 	& \targc  	&\qwc		&\targc	&\qwc		&\qwc		&\targc	&\qwc		&\qwc		&\qwc		&\qwc		&\qwc		& \rstick{y_3}    \\
\lstick{\ket{0}} 	& \qwc	& \targc	& \qwc  	&\ctrlc{-2}	&\qwc		&\qwc		&\targc	&\qwc		&\qwc		&\qwc		&\qwc		&\qwc		&\qwc		& \rstick{y_2}    \\
\lstick{\ket{0}}	& \targc 	& \ctrlc{-1}	& \ctrlc{-2}  	&\qwc		&\qwc		&\targc	&\qwc		&\qwc		&\qwc		&\qwc		&\qwc		&\qwc		&\qwc		& \rstick{y_1}       
}}}
}{%
  \fcaption{$S_{27}$ quantum circuit.}\label{S27circuit}%
}
\end{floatrow}
% end circuits
%
\vspace{20pt}
%
% start tables
\begin{floatrow}
\capbtabbox{%
\scalebox{0.85}{
\begin{tabular}{l l l l l | r r r r r }  
$x_5$ & $x_4$ & $x_3$ & $x_2$ & $x_1$ & $y_5$ & $y_4$ & $y_3$ & $y_2$ & $y_1$ \\ 
  \hline \hline
   0 & 0 & 0 & 0 & 0 & 0 & 0 & 0 & 0 & 0 \\
   0 & 0 & 0 & 0 & 1 & 0 & 0 & 0 & 0 & 1 \\
   0 & 0 & 0 & 1 & 0 & 0 & 0 & 0 & 1 & 0 \\
   0 & 0 & 0 & 1 & 1 & 0 & 0 & 0 & 1 & 1 \\
   0 & 0 & 1 & 0 & 0 & 0 & 0 & 1 & 0 & 0 \\
   0 & 0 & 1 & 0 & 1 & 0 & 0 & 1 & 0 & 1 \\
   0 & 0 & 1 & 1 & 0 & 0 & 0 & 1 & 1 & 0 \\
   0 & 0 & 1 & 1 & 1 & 0 & 0 & 1 & 1 & 1 \\
   0 & 1 & 0 & 0 & 0 & 1 & 1 & 0 & 0 & 0 \\ 
   0 & 1 & 0 & 0 & 1 & 1 & 1 & 0 & 0 & 1 \\ 
   0 & 1 & 0 & 1 & 0 & 1 & 1 & 0 & 1 & 0 \\ 
   0 & 1 & 0 & 1 & 1 & 1 & 1 & 0 & 1 & 1 \\
   0 & 1 & 1 & 0 & 0 & 1 & 1 & 1 & 0 & 0 \\ 
   0 & 1 & 1 & 0 & 1 & 1 & 1 & 1 & 0 & 1 \\
   0 & 1 & 1 & 1 & 0 & 1 & 1 & 1 & 1 & 0 \\ 
   0 & 1 & 1 & 1 & 1 & 1 & 1 & 1 & 1 & 1 \\
   1 & 0 & 0 & 0 & 0 & 1 & 0 & 0 & 0 & 0 \\ 
   1 & 0 & 0 & 0 & 1 & 1 & 0 & 0 & 0 & 1 \\
   1 & 0 & 0 & 1 & 0 & 1 & 0 & 0 & 1 & 0 \\ 
   1 & 0 & 0 & 1 & 1 & 1 & 0 & 0 & 1 & 1 \\
   1 & 0 & 1 & 0 & 0 & 1 & 0 & 1 & 0 & 0 \\ 
   1 & 0 & 1 & 0 & 1 & 1 & 0 & 1 & 0 & 1 \\
   1 & 0 & 1 & 1 & 0 & 1 & 0 & 1 & 1 & 0 \\ 
   1 & 0 & 1 & 1 & 1 & 1 & 0 & 1 & 1 & 1 \\
   1 & 1 & 0 & 0 & 0 & 0 & \underline{\underline{1}} & 0 & \underline{1} & \underline{1} \\ \hline
   1 & 1 & 0 & 0 & 1 & 0 & \underline{0} & 0 & 0 & \underline{0} \\ 
   1 & 1 & 0 & 1 & 0 & 0 & \underline{0} & 0 & \underline{0} & \underline{1} \\ 
   1 & 1 & 0 & 1 & 1 & 0 & \underline{0} & 0 & 1 & \underline{0} \\
   1 & 1 & 1 & 0 & 0 & 0 & \underline{\underline{\underline{0}}} & \underline{0} & \underline{1} & \underline{1} \\ 
   1 & 1 & 1 & 0 & 1 & 0 & \underline{0} & 1 & 0 & \underline{0} \\
   1 & 1 & 1 & 1 & 0 & 0 & \underline{0} & 1 & \underline{0} & \underline{1} \\ 
   1 & 1 & 1 & 1 & 1 & 0 & \underline{0} & 1 & 1 & \underline{0} \\
\end{tabular}
}
}{%
  \tcaption{$S_{25}$ truth table.}\label{S25table}%
}
\capbtabbox{%
\scalebox{0.85}{
\begin{tabular}{l l l l l | r r r r r }  
$x_5$ & $x_4$ & $x_3$ & $x_2$ & $x_1$ & $y_5$ & $y_4$ & $y_3$ & $y_2$ & $y_1$ \\ 
  \hline \hline
   0 & 0 & 0 & 0 & 0 & 0 & 0 & 0 & 0 & 0 \\
   0 & 0 & 0 & 0 & 1 & 0 & 0 & 0 & 0 & 1 \\
   0 & 0 & 0 & 1 & 0 & 0 & 0 & 0 & 1 & 0 \\
   0 & 0 & 0 & 1 & 1 & 0 & 0 & 0 & 1 & 1 \\
   0 & 0 & 1 & 0 & 0 & 0 & 1 & 1 & 0 & 0 \\
   0 & 0 & 1 & 0 & 1 & 0 & 1 & 1 & 0 & 1 \\
   0 & 0 & 1 & 1 & 0 & 0 & 1 & 1 & 1 & 0 \\
   0 & 0 & 1 & 1 & 1 & 0 & 1 & 1 & 1 & 1 \\
   0 & 1 & 0 & 0 & 0 & 1 & 1 & 0 & 0 & 0 \\ 
   0 & 1 & 0 & 0 & 1 & 1 & 1 & 0 & 0 & 1 \\ 
   0 & 1 & 0 & 1 & 0 & 1 & 1 & 0 & 1 & 0 \\ 
   0 & 1 & 0 & 1 & 1 & 1 & 1 & 0 & 1 & 1 \\
   0 & 1 & 1 & 0 & 0 & 1 & 0 & 1 & 0 & 0 \\ 
   0 & 1 & 1 & 0 & 1 & 1 & 0 & 1 & 0 & 1 \\
   0 & 1 & 1 & 1 & 0 & 1 & 0 & 1 & 1 & 0 \\ 
   0 & 1 & 1 & 1 & 1 & 1 & 0 & 1 & 1 & 1 \\
   1 & 0 & 0 & 0 & 0 & 1 & 0 & 0 & 0 & 0 \\ 
   1 & 0 & 0 & 0 & 1 & 1 & 0 & 0 & 0 & 1 \\
   1 & 0 & 0 & 1 & 0 & 1 & 0 & 0 & 1 & 0 \\ 
   1 & 0 & 0 & 1 & 1 & 1 & 0 & 0 & 1 & 1 \\
   1 & 0 & 1 & 0 & 0 & 1 & 1 & 1 & 0 & 0 \\ 
   1 & 0 & 1 & 0 & 1 & 1 & 1 & 1 & 0 & 1 \\
   1 & 0 & 1 & 1 & 0 & 1 & 1 & 1 & 1 & 0 \\ 
   1 & 0 & 1 & 1 & 1 & 1 & 1 & 1 & 1 & 1 \\
   1 & 1 & 0 & 0 & 0 & 0 & 1 & 0 & 0 & \underline{1} \\ 
   1 & 1 & 0 & 0 & 1 & 0 & 1 & 0 & \underline{1} & \underline{0} \\ 
   1 & 1 & 0 & 1 & 0 & 0 & 1 & 0 & 1 & \underline{1} \\ \hline
   1 & 1 & 0 & 1 & 1 & 0 & \underline{0} & 0 & \underline{0} & \underline{0} \\
   1 & 1 & 1 & 0 & 0 & 0 & 0 & \underline{0} & 0 & \underline{1} \\ 
   1 & 1 & 1 & 0 & 1 & 0 & 0 & \underline{0} & \underline{1} & \underline{0} \\
   1 & 1 & 1 & 1 & 0 & 0 & 0 & \underline{0} & 1 & \underline{1} \\ 
   1 & 1 & 1 & 1 & 1 & 0 & \underline{1} & \underline{\underline{1}} & \underline{0} & \underline{0} \\
\end{tabular}
}
}{%
  \tcaption{$S_{27}$ truth table.}\label{S27table}%
}
\end{floatrow}
\end{figure}
% end S_25 and S_27

%% S_29 and S_31
\begin{figure}[H]
\begin{floatrow}
\ffigbox{%
\scalebox{0.85}{\centerline{
\Qcircuit @C=.7em @R=.7em { 
\lstick{x_5} 		&\ctrl{1}	& \qw		&\qw  		&\qw		&\qw		&\qw		&\qw		&\ctrl{6}	&\qw		&\qw		&\ctrl{5}	&\qw		& \rstick{x_5}    \\
\lstick{x_4} 		&\ctrl{8}	& \qw		&\qw  		&\qw		&\qw		&\qw		&\qw		&\qw		&\qw		&\ctrl{4}	&\qw		&\qw		& \rstick{x_4}    \\
\lstick{x_3} 		&\qw		& \ctrl{5}	&\qw  		&\qw		&\qw		&\qw		&\ctrl{5}	&\qw		&\ctrl{4}	&\qw		&\qw		&\qw		& \rstick{x_3}    \\
\lstick{x_2} 		&\qw		& \qw 		& \qw  	&\ctrlo{4}	&\qw		&\ctrl{5}	&\qw		&\qw		&\qw		&\qw		&\qw		&\qw		& \rstick{x_2}    \\
\lstick{x_1}	 	& \qw 		& \qw		& \ctrlo{3}  	&\qw		&\ctrl{5}	&\qw		&\qw		&\qw		&\qw		&\qw		&\qw		&\qw		& \rstick{x_1}    \\
\lstick{\ket{0}} 	& \qwc	& \qwc 	& \qwc  	&\qwc		&\qwc		&\qwc		&\qwc		&\qwc		&\qwc		&\targc	&\targc	&\qwc		& \rstick{y_5}    \\
\lstick{\ket{0}} 	& \qwc	& \qwc 	& \qwc  	&\qwc		&\qwc		&\qwc		&\qwc		&\targc	&\targc	&\qwc		&\qwc		&\qwc		& \rstick{y_4}    \\
\lstick{\ket{0}} 	& \qwc	& \targc 	& \ctrlc{1}  	&\targc	&\qwc		&\qwc		&\targc	&\qwc		&\qwc		&\qwc		&\qwc		&\qwc		& \rstick{y_3}    \\
\lstick{\ket{0}} 	& \qwc	& \qwc	& \targc  	&\ctrlc{-1}	&\qwc		&\targc	&\qwc		&\qwc		&\qwc		&\qwc		&\qwc		&\qwc		& \rstick{y_2}    \\
\lstick{\ket{0}}	& \targc 	& \ctrlc{-2}	& \qwc  	&\qwc		&\targc	&\qwc		&\qwc		&\qwc		&\qwc		&\qwc		&\qwc		&\qwc		& \rstick{y_1}       
}}}
}{%
  \fcaption{$S_{29}$ quantum circuit.}\label{S29circuit}%
}
\ffigbox{%
\scalebox{0.85}{\centerline{
\Qcircuit @C=.7em @R=.7em { 
\lstick{x_5} 		&\qw		& \qw		& \qw 		& \ctrl{1}  	&\qw		&\qw		&\qw		&\qw		&\qw		&\ctrl{5}	&\qw		& \rstick{x_5}    \\
\lstick{x_4} 		&\qw		& \qw 		& \qw 		& \ctrl{8}  	&\qw		&\qw		&\qw		&\qw		&\ctrl{5}	&\qw		&\qw		& \rstick{x_4}    \\
\lstick{x_3}	 	&\qw 		& \qw		& \ctrl{5} 	& \qw  	&\qw		&\qw		&\ctrl{4}	&\qw		&\qw		&\qw		&\qw		& \rstick{x_3}    \\
\lstick{x_2} 		&\qw		& \ctrl{5} 	& \qw 		& \qw  	&\qw		&\ctrl{4}	&\qw		&\qw		&\qw		&\qw		&\qw		& \rstick{x_2}    \\
\lstick{x_1}	 	& \ctrl{5} 	& \qw		& \qw 		& \qw  	&\ctrl{4}	&\qw		&\qw		&\qw		&\qw		&\qw		&\qw		& \rstick{x_1}    \\
\lstick{\ket{0}} 	& \qwc	& \qwc 	& \qwc 	& \qwc  	&\qwc		&\qwc		&\qwc		&\targc	&\qwc		&\targc	&\qwc		& \rstick{y_5}    \\
\lstick{\ket{0}} 	& \qwc	& \qwc 	& \qwc 	& \qwc  	&\qwc		&\qwc		&\targc	&\ctrlc{-1}	&\targc	&\qwc		&\qwc		& \rstick{y_4}    \\
\lstick{\ket{0}} 	& \qwc	& \qwc 	& \targc 	& \qwc  	&\qwc		&\targc	&\ctrloc{-1}	&\qwc		&\qwc		&\qwc		&\qwc		& \rstick{y_3}    \\
\lstick{\ket{0}} 	& \qwc	& \targc  	& \qwc 	& \qwc 	&\targc	&\ctrloc{-1}	&\qwc		&\qwc		&\qwc		&\qwc		&\qwc		& \rstick{y_2}    \\
\lstick{\ket{0}}	& \targc 	& \qwc	& \qwc 	& \targc  	&\ctrloc{-1}	&\qwc		&\qwc		&\qwc		&\qwc		&\qwc		&\qwc		& \rstick{y_1}         
}}}
}{%
  \fcaption{$S_{31}$ quantum circuit.}\label{S31circuit}%
}
\end{floatrow}
% end circuits
%
\vspace{20pt}
%
% start tables
\begin{floatrow}
\capbtabbox{%
\scalebox{0.85}{
\begin{tabular}{l l l l l | r r r r r }  
$x_5$ & $x_4$ & $x_3$ & $x_2$ & $x_1$ & $y_5$ & $y_4$ & $y_3$ & $y_2$ & $y_1$ \\ 
  \hline \hline
   0 & 0 & 0 & 0 & 0 & 0 & 0 & 0 & 0 & 0 \\
   0 & 0 & 0 & 0 & 1 & 0 & 0 & 0 & 0 & 1 \\
   0 & 0 & 0 & 1 & 0 & 0 & 0 & 0 & 1 & 0 \\
   0 & 0 & 0 & 1 & 1 & 0 & 0 & 0 & 1 & 1 \\
   0 & 0 & 1 & 0 & 0 & 0 & 1 & 1 & 0 & 0 \\
   0 & 0 & 1 & 0 & 1 & 0 & 1 & 1 & 0 & 1 \\
   0 & 0 & 1 & 1 & 0 & 0 & 1 & 1 & 1 & 0 \\
   0 & 0 & 1 & 1 & 1 & 0 & 1 & 1 & 1 & 1 \\
   0 & 1 & 0 & 0 & 0 & 1 & 0 & 0 & 0 & 0 \\ 
   0 & 1 & 0 & 0 & 1 & 1 & 0 & 0 & 0 & 1 \\ 
   0 & 1 & 0 & 1 & 0 & 1 & 0 & 0 & 1 & 0 \\ 
   0 & 1 & 0 & 1 & 1 & 1 & 0 & 0 & 1 & 1 \\
   0 & 1 & 1 & 0 & 0 & 1 & 1 & 1 & 0 & 0 \\ 
   0 & 1 & 1 & 0 & 1 & 1 & 1 & 1 & 0 & 1 \\
   0 & 1 & 1 & 1 & 0 & 1 & 1 & 1 & 1 & 0 \\ 
   0 & 1 & 1 & 1 & 1 & 1 & 1 & 1 & 1 & 1 \\
   1 & 0 & 0 & 0 & 0 & 1 & 1 & 0 & 0 & 0 \\ 
   1 & 0 & 0 & 0 & 1 & 1 & 1 & 0 & 0 & 1 \\
   1 & 0 & 0 & 1 & 0 & 1 & 1 & 0 & 1 & 0 \\ 
   1 & 0 & 0 & 1 & 1 & 1 & 1 & 0 & 1 & 1 \\
   1 & 0 & 1 & 0 & 0 & 1 & 0 & 1 & 0 & 0 \\ 
   1 & 0 & 1 & 0 & 1 & 1 & 0 & 1 & 0 & 1 \\
   1 & 0 & 1 & 1 & 0 & 1 & 0 & 1 & 1 & 0 \\ 
   1 & 0 & 1 & 1 & 1 & 1 & 0 & 1 & 1 & 1 \\
   1 & 1 & 0 & 0 & 0 & 0 & 1 & 0 & 0 & \underline{1} \\ 
   1 & 1 & 0 & 0 & 1 & 0 & 1 & 0 & 0 & \underline{0} \\ 
   1 & 1 & 0 & 1 & 0 & 0 & 1 & 0 & 1 & \underline{1} \\ 
   1 & 1 & 0 & 1 & 1 & 0 & 1 & 0 & 1 & \underline{0} \\
   1 & 1 & 1 & 0 & 0 & 0 & 0 & \underline{\underline{1}} & \underline{1} & \underline{1} \\ \hline
   1 & 1 & 1 & 0 & 1 & 0 & 0 & \underline{0} & 0 & \underline{0} \\
   1 & 1 & 1 & 1 & 0 & 0 & 0 & \underline{0} & \underline{0} & \underline{1} \\ 
   1 & 1 & 1 & 1 & 1 & 0 & 0 & \underline{0} & 1 & \underline{0} \\
\end{tabular}
}
}{%
  \tcaption{$S_{29}$ truth table.}\label{S29table}%
}
\capbtabbox{%
\scalebox{0.85}{
\begin{tabular}{l l l l l | r r r r r }  
$x_5$ & $x_4$ & $x_3$ & $x_2$ & $x_1$ & $y_5$ & $y_4$ & $y_3$ & $y_2$ & $y_1$ \\ 
  \hline \hline
   0 & 0 & 0 & 0 & 0 & 0 & 0 & 0 & 0 & 0 \\
   0 & 0 & 0 & 0 & 1 & 0 & 0 & 0 & 0 & 1 \\
   0 & 0 & 0 & 1 & 0 & 0 & 0 & 0 & 1 & 0 \\
   0 & 0 & 0 & 1 & 1 & 0 & 0 & 0 & 1 & 1 \\
   0 & 0 & 1 & 0 & 0 & 0 & 0 & 1 & 0 & 0 \\
   0 & 0 & 1 & 0 & 1 & 0 & 0 & 1 & 0 & 1 \\
   0 & 0 & 1 & 1 & 0 & 0 & 0 & 1 & 1 & 0 \\
   0 & 0 & 1 & 1 & 1 & 0 & 0 & 1 & 1 & 1 \\
   0 & 1 & 0 & 0 & 0 & 0 & 1 & 0 & 0 & 0 \\ 
   0 & 1 & 0 & 0 & 1 & 0 & 1 & 0 & 0 & 1 \\ 
   0 & 1 & 0 & 1 & 0 & 0 & 1 & 0 & 1 & 0 \\ 
   0 & 1 & 0 & 1 & 1 & 0 & 1 & 0 & 1 & 1 \\
   0 & 1 & 1 & 0 & 0 & 0 & 1 & 1 & 0 & 0 \\ 
   0 & 1 & 1 & 0 & 1 & 0 & 1 & 1 & 0 & 1 \\
   0 & 1 & 1 & 1 & 0 & 0 & 1 & 1 & 1 & 0 \\ 
   0 & 1 & 1 & 1 & 1 & 0 & 1 & 1 & 1 & 1 \\
   1 & 0 & 0 & 0 & 0 & 1 & 0 & 0 & 0 & 0 \\ 
   1 & 0 & 0 & 0 & 1 & 1 & 0 & 0 & 0 & 1 \\
   1 & 0 & 0 & 1 & 0 & 1 & 0 & 0 & 1 & 0 \\ 
   1 & 0 & 0 & 1 & 1 & 1 & 0 & 0 & 1 & 1 \\
   1 & 0 & 1 & 0 & 0 & 1 & 0 & 1 & 0 & 0 \\ 
   1 & 0 & 1 & 0 & 1 & 1 & 0 & 1 & 0 & 1 \\
   1 & 0 & 1 & 1 & 0 & 1 & 0 & 1 & 1 & 0 \\ 
   1 & 0 & 1 & 1 & 1 & 1 & 0 & 1 & 1 & 1 \\
   1 & 1 & 0 & 0 & 0 & 1 & 1 & 0 & 0 & \underline{1} \\
   1 & 1 & 0 & 0 & 1 & 1 & 1 & 0 & \underline{1} & \underline{0} \\ 
   1 & 1 & 0 & 1 & 0 & 1 & 1 & 0 & 1 & \underline{1} \\ 
   1 & 1 & 0 & 1 & 1 & 1 & 1 & \underline{1} & \underline{0} & \underline{0} \\
   1 & 1 & 1 & 0 & 0 & 1 & 1 & 1 & 0 & \underline{1} \\ 
   1 & 1 & 1 & 0 & 1 & 1 & 1 & 1 & \underline{1} & \underline{0} \\
   1 & 1 & 1 & 1 & 0 & 1 & 1 & 1 & 1 & \underline{1} \\  \hline
   1 & 1 & 1 & 1 & 1 & \underline{0} & \underline{0} & \underline{0} & \underline{0} & \underline{0} \\
\end{tabular}
}
}{%
  \tcaption{$S_{31}$ truth table.}\label{S31table}%
}
\end{floatrow}
\end{figure}
% end S_29 and S_31

\nonumsection{References}
\noindent

%\bibliographystyle{unsrt}
%\bibliography{periodicBib}

\end{document}